\documentstyle[11pt,amsfonts,fullpage]{amsart}

\newtheorem{theorem}{Theorem}[section]

\newtheorem{remark}[theorem]{Remark}

\begin{document}

\title[Space-Times in $3+1$ and $4+1$ Dimensions]{Remarks on Evolution of 
Space-Times in 3+1 and 4+1 Dimensions}

\author[M.T. Anderson]{Michael T. Anderson}

\maketitle

\abstract
A large class of vacuum space-times is constructed in dimension 4+1 
from hyperboloidal initial data sets which are not small perturbations 
of empty space data. These space-times are future geodesically 
complete, smooth up to their future null infinity ${\cal I}^{+}$, and 
extend as vacuum space-times through their Cauchy horizon. Dimensional 
reduction gives non-vacuum space-times with the same properties in 3+1 
dimensions.

\medskip


\endabstract

\setcounter{section}{1}
\setcounter{equation}{0}

\medskip

{\bf \S 1. Introduction}

\medskip

 Let $(\Sigma, g)$ be a complete Riemannian Einstein $n$-manifold with 
negative scalar curvature, normalized so that
\begin{equation} \label{e1.1}
Ric_{g} = -(n-1)g. 
\end{equation}
It is well-known that the Lorentzian cone over $\Sigma$, i.e. the 
metric
\begin{equation} \label{e1.2}
{\bf g} = - d\tau^{2} + \tau^{2}g 
\end{equation}
on {\bf M} = ${\Bbb R}^{+} \times \Sigma $ is a vacuum solution to the 
Einstein equations in $n+1 = (1,n)$ dimensions. In the case of 3+1 
dimensions, all Einstein metrics $g$ on a 3-manifold $\Sigma $ are of 
constant curvature and hence the 4-metric {\bf g} is flat; in fact 
({\bf M, g}) is then just (a quotient of) the interior of the future 
light cone of a point in empty Minkowski space. In dimensions higher 
than 3, solutions of (1.1) are usually not of constant curvature and 
hence the space-times {\bf (M, g)} are typically not flat.

 For $(\Sigma, g)$ as in (1.1), the space-time ({\bf M, g}) in (1.2) 
is globally hyperbolic with Cauchy surface given by the $n$-manifold 
$\Sigma$. The Cauchy data on $\Sigma$ are of the form
\begin{equation} \label{e1.3}
(g, K) = (g, g), 
\end{equation}
where $K$ is the extrinsic curvature. This Cauchy data on $\Sigma $ 
forms a hyperboloidal initial data set, c.f. [11], [3]. An important 
result of Friedrich [11] implies that on a 3-dimensional hyperboloidal 
initial data set $\Sigma $ which is a small, suitable perturbation of 
standard hyperboloidal data in Minkowski space-time, one has long-time 
evolution to the future of $\Sigma$, with a smooth future null infinity 
${\cal I}^{+}$. However, one does not expect long-time future evolution 
to hold for large perturbations off the standard hyperboloidal initial 
data in Minkowski space, in that one expects formation of black holes 
and singularities to the finite proper time future of $\Sigma$, 
at least generically.

 The first purpose of this paper is to point out that such reasoning 
or expected behavior may not hold in higher, in particular $4+1$ dimensions. 
Thus, a large class of hyperboloidal initial data are constructed in $4+1$
dimensions for which the vacuum space-time {\bf (M, g)} in (1.2) is future 
geodesically complete with smooth future null infinity ${\cal I}^{+}$, 
and which may be considered as {\it large} perturbations of the standard 
flat data, c.f. Theorem 3.1. Moreover, if Friedrich's theorem [11] 
generalizes to higher dimensions, one would then have {\it open sets} 
of such large initial data whose maximal Cauchy developments have the 
stated properties. This indicates there may be some strong differences 
between vacuum evolution of space-times in $3+1$ and higher dimensions. 

 A second purpose is to study the behavior of the space-times (1.2) at 
the Cauchy horizon $\Sigma_{0}$ where $\tau = 0$. In Theorem 4.2, we 
prove that the space-times {\bf (M, g)} extend as $C^{1, 1/2}$ weak 
solutions of the vacuum Einstein equations across the horizon 
$\Sigma_{0}$. This gives a large class of space-times in $4+1$ dimensions, 
depending in fact on 2 free functions of 3 variables, which have 
extensions past a Cauchy horizon. The behavior of the curvature tensor 
{\bf R} of {\bf (M, g)} at $\Sigma_{0}$ is singular, although the 
singularity is relatively mild. The curvature blows up at most as 
$x^{-1/2}$, where $x$ is the Lorentzian distance to $\Sigma_{0}$. If 
one allows such singularities, then this class of space-times does 
not satisfy the strong cosmic censorship hypothesis in this dimension. 

\medskip

  The results above point out some differences in the evolution of vacuum 
space-times in 3+1 and higher dimensions. However, if the initial data 
(1.1) on a 4-manifold $\Sigma$ has an isometric $S^{1}$ action, then 
the vacuum space-time {\bf (M, g)} can be dimensionally reduced to a 
3+1 dimensional space-time, with Cauchy surface $\Sigma / S^{1}$. 
The resulting $3+1$ space-times have matter terms given by a massless 
scalar field or more generally by a wave map (non-linear $\sigma$-model) 
with target the hyperbolic plane $H^{2}(-1)$. The results above on 
long-time future evolution, smoothness of ${\cal I}^{+}$, and extension 
past Cauchy horizons also hold for such $3+1$ space-times. This situation 
is analysed in \S 5.

\setcounter{section}{2}
\setcounter{equation}{0}

\medskip

{\bf \S 2. Initial data set and space-time (M, g)}

\medskip

 Let $\Sigma $ be a compact manifold with non-empty boundary 
$\partial\Sigma .$ A complete Riemannian metric $g$ on $\Sigma $ is 
conformally compact if there is a defining function $\rho $ for 
$\partial\Sigma $ on $\Sigma $ such that the conformally equivalent 
metric
\begin{equation} \label{e2.1}
\bar g = \rho^{2}g 
\end{equation}
extends to a metric on the closure $\bar \Sigma = 
\Sigma\cup\partial\Sigma .$ The compactification $\bar g$ is not 
uniquely determined by $g$, since there are many choices for the 
defining function. However, the conformal class $[\bar g]$ is 
well-defined. Given $\rho ,$ the metric $\gamma  = \bar 
g|_{\partial\Sigma}$ is called the boundary metric associated to $g$ 
and $\rho $ and the conformal class $[\gamma ]$ on $\partial\Sigma $ is 
the conformal infinity of $(\Sigma , g)$ The metric $g$ is called 
$C^{m,\alpha}$ conformally compact if there is a compactification $\bar 
g$ as in (2.1) which extends to a $C^{m,\alpha}$ metric on $\bar 
\Sigma$; here $C^{m, \alpha}$ is the class of metrics which are $C^{m}$ 
with $\alpha$-H\"older continuous $m$th derivatives in some (e.g. 
harmonic) coordinate system.

 We begin by quoting the following result from [2]. Let $Met^{0}(S^{3}) 
= Met^{0}_{m,\alpha}(S^{3})$ be the component of the space of metrics of 
non-negative scalar curvature on $S^{3}$ containing the round metric, 
and let ${\cal C}^{0} = {\cal C}^{0}_{m, \alpha}$ be the space of 
conformal classes of $C^{m, \alpha}$ metrics on $S^{3}$ containing a 
representative in $Met^{0}(S^{3}).$ Let $B^{4}$ be the 4-ball, so that 
$\partial B^{4} = S^{3}$.
\begin{theorem} \label{t 2.1.}
  Let $\gamma $ be any $C^{m+1,\alpha}$ metric in $Met^{0}_{m+1, 
\alpha}(S^{3})$, $m \geq  2$. Then there exists a complete 
$C^{m+1,\alpha}$ conformally compact Einstein metric $g = g_{\gamma}$ 
on $\Sigma  = B^{4}$ with conformal infinity $[\gamma]$.

 This result also holds when $m = \infty$, and $m = \omega$, giving 
$C^{\infty}$, respectively real-analytic, ($C^{\omega}$), conformally 
compact Einstein metrics with prescribed conformal infinity 
$[\gamma] \in {\cal C}^{0}$.
\end{theorem}

 Given a choice of boundary metric $\gamma\in [\gamma ],$ within a 
neighborhood $U$ of $\partial\Sigma = S^{3}$ the metric $g$ may be 
written in the form
\begin{equation} \label{e2.2}
g = \frac{1}{1+s^{2}}ds^{2} + s^{2}\gamma  + h, 
\end{equation}
where the parameter $s \in  [s_{o}, \infty ),$ for some $s_{o}\in{\Bbb 
R} $ and the bilinear form $h$ is bounded (w.r.t. $g$). Under the 
substitution $s = sinhr$, one has 
\begin{equation} \label{e2.3}
(1+s^{2})^{-1}ds^{2} + s^{2}\gamma  = dr^{2} + sinh^{2}r\gamma , 
\end{equation}
so that this part of the metric is a hyperbolic cone metric over the 
boundary metric $\gamma .$ Of course the truncated cone metric in (2.3) 
is not Einstein unless $\gamma  = \gamma_{o},$ the round metric on 
$S^{3}.$ The term $h$ is a perturbation term, possibly large into the 
interior, and Theorem 2.1 implies that $h$ may chosen so that $g$ in 
(2.2) satisfies the Einstein equations (1.1) globally. The function 
\begin{equation} \label{e2.4}
\rho  = 2e^{-r} = 2e^{-(arcsinh s)} \sim  2/s 
\end{equation}
is a geodesic defining function for $\partial\Sigma $ in $(\Sigma, g)$, 
in that the integral curves of $\nabla r$ on $(\Sigma, g)$ are 
geodesics. Similarly, the compactification 
\begin{equation} \label{e2.5}
\bar g = \rho^{2}g 
\end{equation}
is a geodesic compactification, in that $\rho (x) = dist_{\bar g}(x, 
\partial\Sigma )$, with boundary metric $\gamma$. This compactification 
may not have the optimal regularity, but for $\gamma $ as in Theorem 
2.1, $\bar g$ is at least a $C^{m,\alpha}$ conformal compactification. 
Combining (2.2) and (2.4) gives
\begin{equation} \label{e2.6}
\bar g = \rho^{2}g = d\rho^{2} + (1-\frac{1}{4}\rho^{2})\gamma  + 
\rho^{2}h, 
\end{equation}
so that
\begin{equation} \label{e2.7}
\rho^{2}h \in  C^{m,\alpha}(\bar \Sigma), 
\end{equation}
i.e. $\rho^{2}h$ extends $C^{m,\alpha}$ smoothly up to $\partial\Sigma 
.$ Of course since $h$ is bounded, $\rho^{2}h =$ 0 on $\partial\Sigma 
.$ The coordinate expressions (2.2) or (2.6) are valid up to the cut 
locus of $\partial\Sigma $ in $(\Sigma , \bar g).$ One sees that the 
curvature tensor $R_{g}$ of $g$ in (2.2) approaches the curvature 
tensor $R_{-1}$ of the hyperbolic metric at a rate
\begin{equation} \label{e2.8}
|R_{g} -  R_{-1}| = O(\rho^{2}) = O(s^{-2}), 
\end{equation}
but not any faster in general. Thus, such metrics $g$ are 
asymptotically hyperbolic.

\medskip

  As noted in the Introduction, given a conformally compact Einstein 
4-manifold $(\Sigma, g)$ as above, the space-time {\bf (M, g)} in (1.2) 
is globally hyperbolic with Cauchy surface $\Sigma$ and Cauchy data 
given by (1.3). Observe that Theorem 2.1 provides a large space of 
such initial data. Namely, the collection of such metrics $g$ on the 
$4$-ball $\Sigma = B^{4}$ are effectively parametrized by the space of 
their conformal infinities, i.e. the boundary conformal classes 
$[\gamma]\in{\cal C}^{0}.$ The space ${\cal C}^{0}$ is a ``large'' 
space, in that for instance there are curves of unit volume metrics 
in ${\cal C}^{0}$ starting at the standard round metric $\gamma_{0}$ 
and diverging arbitrarily far away from $\gamma_{0}$ in terms of the 
size of the curvature or diameter. The resulting Einstein metrics on 
$\Sigma$ are then also arbitrarily far away from the hyperbolic metric. 

 The time parameter $\tau$ varies over the domain $(0, \infty)$ and 
w.r.t. the initial Cauchy surface $\Sigma$ the time evolution of the 
space-time is just given by rescalings of $(\Sigma, g)$; thus, if 
$\Sigma_{\tau}$ is the $\tau$-level set of the function $\tau ,$ then 
${\bf g}|_{\Sigma_{\tau}} = g_{\tau} = \tau^{2}g.$ Of course $\Sigma  = 
\Sigma_{1}.$ The mean curvature $H$ of $\Sigma_{\tau}$ is given by $H = 
4/\tau$, while the extrinsic curvature is pure trace.

 Observe also that the space-time ({\bf M, g}) is geodesically complete, 
(both time-like and null), to the future of $\Sigma$. This is not the 
case to the past of $\Sigma$; the surface $\Sigma_{0} = \{\tau  = 0\}$ 
forms the Cauchy horizon. This will be examined in more detail in 
\S 4 below. From (2.2), the space-time metric {\bf g} may be written 
in the form
\begin{equation} \label{e2.9}
{\bf g} = - d\tau^{2} + \tau^{2}(\frac{1}{1+s^{2}}ds^{2} + s^{2}\gamma 
) + \tau^{2}h, 
\end{equation}
in a neighborhood $U$ of $\partial\Sigma ,$ with $h$ a bilinear form on 
$\Sigma .$

\setcounter{section}{3}
\setcounter{equation}{0}
\setcounter{theorem}{0}

\medskip

{\bf \S 3. Asymptotic flatness}

  In this section, we prove that {\bf (M, g)} is asymptotically flat in 
the sense of Penrose.

\begin{theorem} \label{t 3.1.}
  The space-time ({\bf M, g}) is asymptotically flat to the future of 
$\Sigma ,$ in that ({\bf M, g}) has a smooth conformal compactification
\begin{equation} \label{e3.1}
{\bf \bar g} = \Omega^{2}{\bf g}. 
\end{equation}
More precisely, if $g$ is $C^{m+1,\alpha}$ conformally compact as in 
Theorem 2.1, then ${\bf \bar g}$ is (at least) $C^{m,\alpha},$ i.e. 
${\bf \bar g}$ extends $C^{m,\alpha}$ to the future null infinity 
${\cal I}^{+}$ of ({\bf M, g}).
\end{theorem}

 To prove this result, suppose first that $\gamma  = \gamma_{o},$ so 
that the metric $g$ on $\Sigma $ is the Poincar\'e metric $g_{o}$ on the 
4-ball $B^{4}$ and ({\bf M, g}) is the flat ($4+1$ dimensional) Minkowski 
metric ${\bf g_{o}}$ interior to a future light cone. Following the standard 
compactification of Minkowski space, c.f. [13], let
\begin{equation} \label{e3.2}
\Omega^{2} = \frac{4}{4\tau^{2}s^{2}+(1+\tau^{2})^{2}} = 
4cos^{2}\frac{1}{2}(T+R)cos^{2}\frac{1}{2}(T- R), 
\end{equation}
where $\frac{1}{2}(T+R) = arctan \ v$, $\frac{1}{2}(T- R) = arctan \ u$ 
and $(u, v)$ are advanced and retarded null coordinates, given in terms 
of $(\tau , s)$ by $v = \tau (s+(1+s^{2})^{1/2}), u = -\tau (s- 
(1+s^{2})^{1/2}).$ Thus, the compactification ${\bf \bar g_{o}}$ 
interior to the light cone in these coordinates becomes
\begin{equation} \label{e3.3}
{\bf \bar g_{o}}
 = - dT^{2} + dR^{2} + sin^{2}R \gamma_{o}, 
\end{equation}
with $R \geq $ 0 and $T+R\in [0,\pi ], T- R\in [0,\pi ].$ The full 
metric (3.3), without restriction on the range of coordinates, is the 
Einstein static space-time on ${\Bbb R}\times S^{4}.$ 

 Performing exactly the same computation for the conformal 
compactification (3.1) with $\Omega $ as in (3.2)
 gives
\begin{equation} \label{e3.4}
{\bf \bar g}
 = - dT^{2} + dR^{2} + sin^{2}R \gamma  + \Omega^{2}\tau^{2}h, 
\end{equation}
where, as in (2.9), $h$ is defined on $\Sigma $ and so is independent of 
$\tau .$

 Future null infinity ${\cal I}^{+}$ of ({\bf M, g}) is given by the 
locus 
\begin{equation} \label{e3.5}
{\cal I}^{+} = \{T+R = \pi\}, \ {\rm with} \  T\in (\frac{\pi}{2}, \pi 
).  
\end{equation}
It is clear by inspection that the conformal factor $\Omega $ and the 
first part $- dT^{2} + dR^{2} + sin^{2}R \gamma $ of the metric 
${\bf \bar g}$ are $C^{m,\alpha}$ up to ${\cal I}^{+}.$ Further, 
in any region of ({\bf M, g}) where $\tau$ is bounded away from 0 
and $\infty$, corresponding to $T$ in a compact subset of 
$(\frac{\pi}{2}, \pi)$, one has 
$$\Omega^{2}\tau^{2}h \sim  \rho^{2}h,$$ 
(c.f. (2.4)),  which vanishes on ${\cal I}^{+}$ and extends 
$C^{m,\alpha}$ up to ${\cal I}^{+}$ by (2.7). The behavior at 
$\{T = R = \frac{\pi}{2}\}$, corresponding to the null infinity of 
the Cauchy horizon will be discussed in \S 4.

 At future time-like infinity $\iota^{+} = \{T = \pi , R =$ 0\}, one 
also has $\Omega^{2}\tau^{2}h \rightarrow $ 0 so that the space part 
$dR^{2} + sin^{2}R \gamma  + \Omega^{2}\tau^{2}h$ of ${\bf \bar g}
$ approximates the metric cone on $\gamma $ near $\iota^{+}$. This 
completes the proof.
\begin{remark} \label{r 3.2.}
  {\rm As explained in the Introduction, Theorem 3.1 is in quite strong 
contrast to the situation in $3+1$ dimensions where singularities 
are expected to form to the finite future of ``large'' hyperboloidal 
initial data sets, i.e. initial data which are not close to that of
the standard hyperboloid in Minkowski space-time. 

 Theorem 3.1 shows that such singularity formation does not occur, even 
though $(\Sigma, g)$ may be far away from a hyperbolic metric. For 
example, the curvature of {\bf g} near the origin of $\Sigma = \Sigma_{1}$ 
may become arbitrarily large, as $[\gamma]$ varies over ${\cal C}^{0}$. 
While the initial data $(g, g)$ on $\Sigma $ are not generic, in the sense 
of forming an open set in the space of all solutions to the constraint 
equations, the space of initial data is nevertheless large. Namely, it 
is parametrized by ${\cal C}^{0},$ and so includes a ``large'' open set 
of metrics on $\partial\Sigma  = S^{3}$, as described following (2.8). 
Taking into account the equivalences from conformally related metrics 
and diffeomorphisms, the space of initial data formally corresponds 
locally to 2 free functions of 3 variables.

 In fact, it does not seem unreasonable that Friedrich's result [11] can 
be extended to $4+1$ dimensions to include all suitable small perturbations 
of initial data as in (1.3). If so, then this thickening of the initial 
data would give large open sets in the space of initial data having long 
time evolution to the future and smooth ${\cal I}^{+}$.}
\end{remark}

\begin{remark} \label{r 3.3.}
  {\rm The discussion above, in particular Theorems 2.1 and 3.1, also 
hold for other choices of the Cauchy surface $\Sigma $ in place of 
$B^{4}.$ For example, one may choose $\Sigma  = {\Bbb R}^{3}\times S^{1}$, 
with $\partial \Sigma = S^{2} \times S^{1}$, or $\Sigma $ any disc bundle 
over $S^{2}$ of degree $k \geq  2$, so that 
$\partial\Sigma  = S^{3}/{\Bbb Z}_{k}$, c.f. [2]. In the latter case, 
$\Sigma$ does not admit any hyperbolic metric. 

 Similarly, Theorem 3.1 and the discussion above hold for any 
conformally compact Einstein metric (1.1) on a 4-manifold, (or 
$n$-manifold), for example the Euclidean AdS black hole metrics with 
flat or hyperbolic horizons, c.f. [6] and references therein. However, 
a global existence result as in Theorem 2.1 is lacking when the 
boundary metrics are not in ${\cal C}^{0}$, (and lacking altogether in 
dimensions above 4).}
\end{remark}

\setcounter{section}{4}
\setcounter{equation}{0}
\setcounter{theorem}{0}

\medskip

{\bf \S 4. Cauchy horizon}

 Next, we discuss the behavior at the Cauchy horizon $\Sigma_{0} = 
\{\tau  = 0\}$. This was the initial setting of the work of 
Fefferman-Graham [10]. As usual, one must change coordinates near 
$\Sigma_{0},$ and so, following [10], write
\begin{equation} \label{e4.1}
\tau  = v\rho , 
\end{equation}
in a neighborhood of $\tau  = 0$. One could also use the more standard 
coordinate transformation $r = \tau s, t = \tau (1+s^{2})^{1/2}$, but 
this and other coordinate changes lead to equivalent results. Since 
$\rho $ is a geodesic defining function, the compactification $\bar g$ 
in (1.6) splits as
$$\bar g = d\rho^{2} + g_{\rho}, $$
where $g_{\rho}$ is a $C^{m,\alpha}$ curve of metrics on 
$\partial\Sigma $ with $g_{0} = \gamma .$ Hence
\begin{equation} \label{e4.2}
{\bf g} = - d\tau^{2} + \frac{\tau^{2}}{\rho^{2}}(d\rho^{2} + 
g_{\rho}). 
\end{equation}
Substituting the change of variables (4.1) in (4.2), this becomes
$${\bf g} \equiv {\bf g^{+}} = -\rho^{2}dv^{2} - 2\rho vdvd\rho  + 
v^{2}g_{\rho}. $$
The reason for the notation ${\bf g^{+}}$ will become clear later, but 
for now it signifies $\rho  > $ 0 or $\tau  > $ 0. This metric is 
degenerate at $\rho  = 0$, but the degeneracy comes merely from the 
choice of coordinates. Thus, setting $\rho^{2} = x$, the metric {\bf g} 
has the form
\begin{equation} \label{e4.3}
{\bf g} = {\bf g^{+}} = - xdv^{2} - vdvdx + v^{2}g_{\sqrt x}, 
\end{equation}
which is a (non-degenerate) Lorentz metric at $x = 0$ provided $v >  
0$. The Cauchy horizon $\Sigma_{0} = \{\tau  = 0\} = \{x = 0\}$ 
topologically is the cone on $S^{3},$ with null geodesics parametrized 
by $v$ and with (degenerate) metric
$$g_{0} = {\bf g}|_{\Sigma_{0}} = v^{2}\gamma . $$

 Now we examine the smoothness of the metric {\bf g} up to the Cauchy 
horizon, and the possibility of extending the space-time ({\bf M, g}) 
past the horizon. For simplicity, assume that $m = \omega$, i.e. the 
compactification $\bar g$ is real-analytic, (c.f. Theorem 2.1), so that 
the curve $g_{\rho}$ is real-analytic in $\rho$.

 As discussed in [10], the curve of metrics $g_{\rho}$ has a Taylor 
expansion of the form
\begin{equation} \label{e4.4}
g_{\rho} = g_{(0)} + \rho^{2}g_{(2)} + \rho^{3}g_{(3)} + \rho^{4}g_{(4)}+ 
\rho^{5}g_{(5)}+... , 
\end{equation}
where the coefficients $g_{(j)}$ are bilinear forms on $\partial\Sigma 
.$ Note that there is no term linear in $\rho .$ The $g_{(2)}$ term is 
determined locally by the intrinsic geometry of the boundary metric 
$\gamma  = g_{(0)},$ c.f. [10]. 

 However, the term $g_{(3)}$ is global and not determined by the 
boundary metric $\gamma .$ In fact only for very special Einstein 
metrics $(\Sigma , g)$ does one have $g_{(3)} =$ 0. By [1], Einstein 
metrics with $g_{(3)} =$ 0 are critical points for the renormalized 
volume $V$ or action in the sense of the AdS/CFT correspondence, on the 
moduli space of conformally compact Einstein metrics. Further, one has
\begin{equation} \label{e4.5}
g_{(3)} = 0 
\end{equation}
if and only if the expansion (4.4) has only even powers of $\rho .$

\begin{remark} \label{r 4.1.}
  {\rm Certain of the main results of [10] require that the expansion 
(4.4) is in even powers of $\rho .$ While this can always be achieved 
locally, i.e. in a neighborhood of $\partial\Sigma ,$ c.f. [10, Thm. 
2.3], there is at most a finite dimensional space of such conformally 
compact Einstein metrics globally on $\Sigma .$ To see this, the 
evenness of the expansion, (and real-analyticity), implies the 
compactified metric $\bar g$ on $\Sigma $ reflects across 
$\partial\Sigma $ to give a smooth metric $\widetilde g$ on the double 
$\widetilde \Sigma = \Sigma\cup_{\partial\Sigma}\Sigma .$ The manifold 
$\widetilde \Sigma$ is closed, and $\widetilde g$ is a smooth solution 
of the conformally invariant Bach equation, (c.f. [4, Ch.4H]),
$$\delta d(Ric-\frac{s}{6}g) -2W(Ric) = 0. $$
However, on a compact manifold, without boundary, there is at most a 
finite dimensional space of (conformal classes) of solutions to this 
equation, since it is essentially elliptic transverse to conformal 
classes.}
\end{remark}

 Making the substitution $\rho  = \sqrt x$ in (4.4) gives the expansion
\begin{equation} \label{e4.6}
g_{\sqrt x} = g_{(0)} + xg_{(2)} + x\sqrt x g_{(3)} + x^{2}g_{(4)} + 
x^{2}\sqrt x g_{(5)}+... . 
\end{equation}

 It follows that the metric {\bf g} is, in general, at most $C^{1,1/2}$ 
smooth up to the Cauchy horizon $\Sigma_{0}$ in these coordinates. This 
degree of smoothness cannot be improved by changing coordinates. Of 
course if $g_{(3)} = 0$, then {\bf g} is $C^{\omega}$ up to 
$\Sigma_{0}$ by the remark following (4.5).

 On the other hand, we observe that the curvature {\bf R} of ({\bf M, g}) 
is bounded in many directions up to $\Sigma_{0}$, (away from the 
vertex \{0\}). To see this, let $U,V,W$ be vectors tangent to 
$\Sigma_{\tau}$ at a point $(\tau, p)$, $p\in\Sigma_{\tau}$. 
Using the expression (1.2), standard formulas show the curvature 
of ({\bf M, g}) at $(\tau , p)$ is given by 
$${\bf R}(V,W)U = \tau^{-2}[R_{g}(V,W)U -  R_{-1}(V,W)U], $$
while any component of {\bf R} containing a vector orthogonal to 
$\Sigma_{\tau}$ vanishes; here $R_{-1}$ is the curvature of the 
hyperbolic metric $g_{-1}$ on ${\Bbb R}^{4},$ so $R_{-1}(V,W)U = 
g_{-1}(V,U)W -  g_{-1}(W,U)V.$ Under the substitution (4.1), this gives
\begin{equation} \label{e4.7}
{\bf R}(V,W)U = v^{-2}\{\rho^{-2}[R_{g}(V,W)U -  R_{-1}(V,W)U]\}. 
\end{equation}
On approach to any point in $\Sigma_{0}$ away from the vertex \{0\} 
where $v = 0$ we have $\rho  \rightarrow  0$. But by (2.8) the term 
$\rho^{-2}(R_{g}(V,W)U -  R_{-1}(V,W)U)$ remains bounded as $\rho  
\rightarrow $ 0. In fact from standard formulas relating the curvatures 
of $g$ and $\bar g,$ (c.f. [2]), one finds
\begin{equation} \label{e4.8}
lim_{\rho\rightarrow 0}\rho^{-2}[R_{g} -  R_{-1}] = \bar R -  
\frac{1}{2}\bar Ric\wedge\bar g + \frac{\bar s}{3}\bar g \wedge \bar g, 
\end{equation}
where $\bar s, \bar Ric$ and $\bar R$ are the scalar, Ricci and full 
curvature of $\bar g$ in (2.5), evaluated at $\partial\Sigma $ and 
$\wedge $ denotes the Kulkarni-Nomizu product, c.f. [4, Ch.1]. These 
curvatures of $\bar g$ at $\partial \Sigma$ are in fact determined 
by the intrinsic geometry of the boundary metric $\gamma $ 
at $\partial\Sigma$.

 However, the curvature {\bf R} is not bounded in all directions. 
Namely, a simple computation shows that in 2-planes spanned by the 
vector $\partial / \partial v$ and vectors tangent to $S^{3}$, (i.e. 
orthogonal to $\partial / \partial v$ and $\partial / \partial \rho$), 
the curvature blows up as $x^{-1/2}$, where $x$ is the Lorentzian 
distance to $\Sigma_{0}$. (Of course, the curvature blows up at the 
rate of $v^{-2}$ on approach to the singularity \{0\}, unless the 
space-time ({\bf M, g}) is flat).

 Nevertheless, the space-time ({\bf M, g}) extends past the horizon 
$\Sigma $ as a (weak) solution of the vacuum Einstein equations. 
To see this, just extend the variable $x \geq  0$ to allow negative 
values of $x$. It is immediate that the term $- xdv^{2} - vdvdx$ 
in the expression (4.3) for {\bf g} extends smoothly to $x < 0$. 
When $g_{(3)} = 0$, the term $g_{\sqrt x}$ in (4.3) is a power 
series in powers of $x$, so extends real-analytically across 
$\partial\Sigma ,$ giving a real-analytic extension of {\bf g}. When 
$g_{(3)} \neq  0$, the series for $g_{\rho}$ extends analytically in 
$\rho $ to regions where $\rho  <  0$. This corresponds to replacing 
$\sqrt x$, $x > 0$, by $-\sqrt {|x|}$, $x <  0$. Thus, the series for 
$g_{\sqrt x}$ becomes
\begin{equation} \label{e4.9}
g_{\sqrt |x|} = g_{(0)}+ xg_{(2)}-  x\sqrt {|x|} g_{(3)}+ x^{2}g_{(4)}- 
 x^{2}\sqrt {|x|} g_{(5)}+ x^{3}g_{(6)}+ ... , 
\end{equation}
when $x \leq  0$. In terms of $\rho  < $ 0, this has the form
\begin{equation} \label{e4.10}
g_{\rho}^{-} = g_{(0)}-  \rho^{2}g_{(2)}-  \rho^{3}g_{(3)}+ 
\rho^{4}g_{(4)}+\rho^{5}g_{(5)}-  \rho^{6}g_{(6)}+ ... . 
\end{equation}
Reversing the substitution (4.1), i.e. in terms of $(\tau , \rho )$ 
coordinates, the metric {\bf g} $= {\bf g^{+}}$ extends across the 
Cauchy horizon as
\begin{equation} \label{e4.11}
{\bf g^{-}} = d\tau^{2} + \frac{\tau^{2}}{\rho^{2}}(- d\rho^{2} + 
g_{\rho}^{-}). 
\end{equation}
Referring to (4.2), one sees that ${\bf g^{-}}$ is obtained from ${\bf 
g^{+}}$ by interchanging the role of time and space, (in the direction 
$\rho$). Thus, (4.2) formally corresponds to (4.11) under the 
substitution $\tau  \rightarrow  i\tau , \rho  \rightarrow  i\rho ,$ 
with $g_{\rho}^{+} \rightarrow  g_{\rho}^{-}.$ When $g_{(3)} = 0$, one 
sees also that $g_{\rho}^{-} = g_{i\rho}^{+}.$

 The domain of ${\bf g^{-}}$ includes a neighborhood where $\tau  <  
0$, (or $\rho  <  0$). In particular, it follows that this extension of 
({\bf M, g}) to a larger domain extends ${\cal I}^{+}$ to a larger 
domain; in the form (3.5), the parameter $T$ varies over 
$(\frac{\pi}{2}-\delta , \pi ),$ for some $\delta  >  0$.

 We summarize this analysis in the following result.

\begin{theorem} \label{t 4.2.}
  Suppose the boundary metric $\gamma $ on $\partial\Sigma $ is 
real-analytic. Away from the vertex \{0\}, the space-time ({\bf M, g}) 
extends past the Cauchy horizon $\Sigma_{0}$ as a vacuum space-time. 
The extended space-time is $C^{1,1/2}$ at the Cauchy horizon, but is 
$C^{\omega}$ elsewhere. Within its domain, the extended space-time 
is asymptotically flat, and gives a smooth extension of 
${\cal I}^{+}$ to a larger null surface.
\end{theorem}

 We make several remarks on this construction.

\begin{remark} \label{r 4.3.}
  {\rm Consider the construction above with regard to 
the strong cosmic censorship conjecture of Penrose, c.f. [8] and 
references therein. Roughly speaking, this states that a generic (vacuum) 
globally hyperbolic space-time does not admit a space-time extension 
past the Cauchy horizon, i.e. generically singularities form on 
approach to a Cauchy horizon.

 Depending on the exact meaning of singularities, this is not the case 
for the space-times {\bf (M, g)}. While there is a curvature singularity 
at the Cauchy horizon, it is relatively mild. The curvature is integrable 
along curves terminating at $\Sigma_{0}$, and there are no strong 
curvature singularities at $\Sigma_{0}$, c.f. [9]. While the space-times 
{\bf (M, g)} are not generic, in the sense of forming an open set, they do 
comprise a large family of vacuum solutions, c.f. Remark 3.2. 
Similarly, this situation occurs in $4+1$ dimensions, although see \S 5.}
\end{remark}

\begin{remark} \label{r 4.4.}
  {\rm It would be interesting to understand what the maximal (vacuum) 
extension of ({\bf M, g}) is, i.e. how large the domain of ${\bf 
g^{-}}$ is. Note that the hyperbolic cone metric (1.2) can be defined 
equally well in the interior of the past light cone in place of future 
cone. Is it possible that the metric ${\bf g^{-}}$ interpolates 
smoothly between the Cauchy horizons of these isometric past and future 
cones? If so, then of course the vertex $\{0\}$ is a naked singularity.}
\end{remark}

\begin{remark} \label{r 4.5.}
  {\rm We comment briefly on generalizations of the results above to 
higher dimensions $n+1, n >  4$, and to space-times of negative 
cosmological constant. As in Remark 3.3, the results above hold for any 
$C^{m, \alpha}$ conformally compact Einstein metric, on any 
$n$-manifold $\Sigma$, although the part of Theorem 4.2 concerning the 
extension past $\Sigma_{0}$, i.e. the definition of ${\bf g^{-}}$ in (4.11), 
requires real-analyticity.

  First, there is no known analogue of the global result of Theorem 2.1 
in dimension $n >  4$. There are local results in all dimensions, c.f. 
[2], [5], [12], [14], showing that given any conformally compact Einstein 
metric $g_{1}$ on an $n$-manifold $\Sigma$ with conformal infinity 
$[\gamma_{1}],$ there are conformally compact Einstein perturbations 
$g$ of $g_{1},$ essentially parametrized by small perturbations 
$[\gamma ]$ of $[\gamma_{1}].$ However, these lead to just small 
space-time perturbations, and may possibly be covered by the 
perturbation techniques of Friedrich [11], (or more precisely their 
generalization to higher dimensions). On the other hand, in higher 
dimensions one has stronger regularity of the compactification $\bar 
g$ at $\partial\Sigma ;$ in dimension $n+1,$ the expansion (4.4) is 
even up to the $g_{(n-1)}$ term, c.f. [10], [14]. Thus, such metrics 
will be smoother, in particular of bounded curvature, at the Cauchy 
horizon. 

 Of course, one could take the $4+1$ dimensional space-times (1.2) and 
take the product with space-like ${\Bbb R}$ factors. However, this will 
lead to singularities, (of codimension 4), on ${\cal I}^{+}.$}

 {\rm Finally, starting with any Riemannian Einstein metric $(\Sigma , 
g)$ as in Theorem 2.1, (or as above), one may also form "spherical 
cones", giving vacuum space-times with cosmological constant $\Lambda  
= - 6;$ (other values for $\Lambda  < $ 0 may be obtained by 
rescalings). Thus, set
$${\bf g} = - d\tau^{2} + cos^{2} \tau \cdot g,$$ 
defined on {\bf M} = $S^{1}\times \Sigma$, or on the universal cover 
${\Bbb R}\times \Sigma$. Of course, these coordinates break down at the 
Cauchy horizons where $\tau$ is a multiple of $\pm\frac{\pi}{2}$. 
However, the space-time {\bf (M, g)} extends past the Cauchy horizon as 
a vacuum space-time exactly as before.}
\end{remark}

\medskip

{\bf \S 5. Dimensional reduction}

\setcounter{section}{5}
\setcounter{equation}{0}

 Let $(\Sigma, g)$ be a $C^{m, \alpha}$ conformally compact Einstein 
metric, $m \geq 3$, dim $\Sigma = 4$, and suppose the conformal 
infinity $(\partial\Sigma , [\gamma ])$ has a 1-parameter group of 
isometries, so that there is an $S^{1}$ action by isometries. By 
[2, Thm. 3.1], this action extends to an isometric $S^{1}$ action on 
$\Sigma$. This action of course extends to a space-like isometric $S^{1}$ 
action on ({\bf M, g}). Hence, one may dimensionally reduce the 
5-dimensional space-time ({\bf M, g}) to obtain a 4-dimensional 
space-time $({\bf M^{4}, g^{4}}),$ with Cauchy surface $\Sigma^{3} = 
\Sigma^{4} /S^{1}.$ 

 Following the standard dimensional reduction, the 5-metric may be 
written in the form
$${\bf g} = e^{-2\lambda}{\bf g}^{4} + e^{2\lambda}\theta\cdot \theta , 
$$
where $\theta $ is the unit form dual to the Killing field $X$ and 
$\lambda  = log|X|.$ The metric ${\bf g^{4}}$ is a space-time metric on 
the manifold ${\bf M^{4}} = {\Bbb R}^{+}\times \Sigma^{3}$. The field 
equations for ${\bf g^{4}}$ are
\begin{equation} \label{e5.1}
Ric = \frac{1}{2}(4d\lambda\otimes d\lambda  + 
e^{-4\lambda}d\phi\otimes d\phi ),
\end{equation}
\begin{equation} \label{e5.2}
\Delta \lambda  = -\frac{1}{2}e^{-4\lambda}|d\phi|^{2},
\end{equation}
\begin{equation} \label{e5.3}
divd\phi  = 4<d\lambda , d\phi>,
\end{equation}
where $\phi$ is the twist potential, c.f. [15]. These are the vacuum 
equations coupled to wave-map matter, mapping to the hyperbolic plane 
$H^{2}(-1)$, i.e. to a non-linear $\sigma$-model with target $H^{2}(-1)$. 
In particular, if $\phi = const.$, then these are the vacuum equations 
coupled to a massless scalar field. Observe that $Ric \geq  0$, so that 
$({\bf M^{4}, g^{4}})$ satisfies the strong energy condition. 

 The metric ${\bf g^{4}}$ has singularities arising from the fixed 
points of the $S^{1}$ action. We describe one situation, (among many 
possibilities) where there are no singularities. Thus, suppose 
$\Sigma = \Sigma^{4} = {\Bbb R}^{3}\times S^{1},$ so that 
$\partial\Sigma  = S^{2}\times S^{1}$. Remarks 3.3 and 4.5 show 
that the results of \S 2-\S 4 hold for $\Sigma$. Choose any free $S^{1}$ 
action on $S^{2}\times S^{1}$, (there are many), and let $\gamma $ be any 
metric in $Met^{0}(S^{2}\times S^{1})$ invariant under the action.
Hence any conformally compact Einstein metric $g$ on 
${\Bbb R}^{3}\times S^{1}$ with conformal infinity $[\gamma ]$ has 
an isometric $S^{1}$ action. 

 Now observe this action cannot have any fixed points. For all of the 
$S^{1}$ orbits are homotopic in $\Sigma$, since the orbit 
space is connected. The action on the boundary $S^{2}\times 
S^{1}$ is free, so that the orbits are non-trivial in 
$\pi_{1}(\partial\Sigma)$, and hence non-trivial in $\pi_{1}(\Sigma )$. 
But at any fixed point, the orbit is trivial, giving a contradiction.

 It follows that the quotient $\Sigma^{3} = ({\Bbb R}^{3}\times S^{1})/S^{1}$ 
is a smooth 3-manifold, topologically ${\Bbb R}^{3}$. Carrying out the 
construction above gives a smooth space-time $({\bf M^{4}, g^{4}})$ 
with smooth hyperboloidal Cauchy surface $\Sigma_{1} \equiv \Sigma^{3}$. 
As before, this space-time is future geodesically complete, with smooth 
${\cal I}^{+}$ and with time evolution given by rescalings. This gives a 
constant mean curvature foliation $\Sigma_{\tau}$, $\tau \in (0, \infty)$ 
of $({\bf M^{4}, g^{4}})$. Theorem 4.2 also holds for $({\bf M^{4}, g^{4}})$.

 Since the time evolution of the metric ${\bf g^{4}}$ is given by 
rescalings, with time parameter $\tau$ as in (1.2), the potentials 
$\lambda$, $\phi$ evolve, (on the Cauchy surfaces $\Sigma_{\tau}$), 
as $\lambda_{\tau} = \lambda_{1} + log \tau$, 
$\phi_{\tau} = \tau^{4}\phi_{1}$, where $\lambda = \lambda_{1}, 
\phi = \phi_{1}$ are the potentials on $\Sigma = \Sigma_{1}$. We point 
out that if the $S^{1}$ action on the boundary 
$\partial \Sigma^{4} = S^{2} \times S^{1}$ is hypersurface orthogonal, 
i.e. $d\phi = 0$ on $S^{2} \times S^{1}$, then the $S^{1}$ action 
on $\Sigma^{4} = {\Bbb R}^{3} \times S^{1}$ is also hypersurface orthogonal, 
i.e. $d\phi = 0$ on $\Sigma$. This may be seen by using the analogue of 
the field equation (5.3) on the Einstein 4-manifold $(\Sigma, g)$, 
pairing it with $\phi$ and integrating by parts over $(\Sigma, g)$. 
In fact, the data $(\lambda, \phi)$ are essentially determined by 
their boundary values on the conformal infinity $\partial \Sigma$ of 
$\Sigma$.

 Thus again there is large collection of 3-dimensional initial data 
which have global evolution to the future, smooth ${\cal I}^{+}$, and 
which extend past the Cauchy horizon. In this situation, although the 
time evolution is simple, the spatial geometry of the Cauchy surfaces 
$\Sigma_{\tau}$ is complicated in general. By the Riemann mapping 
theorem the conformal infinity $(S^{2}, [\gamma ])$ of $\Sigma^{3}$ 
is unique, but the matter fields $(\lambda , \phi )$ are freely 
specifiable on the boundary, within the constraint of ${\cal C}^{0}$. 
This gives a degree of freedom in the construction corresponding 
locally to 2 free functions of 2 variables. 

  If $\phi = 0$ on the boundary $S^{2}$, then $\phi \equiv 0$ on the 
$3+1$ space-time and the massless scalar field $\lambda$ may be 
specified arbitrarily on the conformal infinity $S^{2}$, subject 
again only to the condition of $\gamma \in {\cal C}^{0}$. Thus, this 
situation is essentially opposite to that considered by Christodoulou 
in his study of the evolution of spherically symmetric $3+1$ gravity 
coupled to a massless scalar field, c.f. [7] and references therein. 

\medskip

{\bf \S 6. Conclusion}

  In conclusion, this paper presents some evidence that central issues in 
classical $3+1$ general relativity, such as singularity and black hole 
formation and the behavior at Cauchy horizons may be special to this 
dimension and analogues of these issues may not hold in higher dimensions. 
If this is the case, then of course the theoretical methods developed 
toward the resolution of these issues must depend on the special nature 
of $3+1$ dimensions.

  This evidence comes from the analysis of the very simple time evolution 
of the much more complicated spatial geometries possible in higher dimensions. 
In this regard, it would be interesting to know if the long time existence 
results of Friedrich [10] generalize to higher dimension. This would 
strengthen the evidence for the differences in $3+1$ and higher dimensions 
considerably.

  In the opposite direction, in the presence of a natural isometry group 
on the conformal infinity of the initial Cauchy surface, the results 
obtained on $4+1$ vacuum space-times are equivalent to those on the 
$3+1$ space-times with matter given by a reasonably general class of 
massless scalar fields or $\sigma$-models to $H^{2}(-1)$.

\medskip
{\bf Acknowlegment} This work was partially supported by 
NSF Grant DMS 0072591. I would like to thank the referees for their 
suggestions in improving the exposition of the article.

\bibliographystyle{plain}

\begin{center}
June 2001 / Oct. 2001
\end{center}
\medskip
\noindent
\address{Department of Mathematics\\
SUNY at Stony Brook\\
Stony Brook, N.Y. 11794-3651\\}
\email{anderson@@math.sunysb.edu}

\end{document}